\documentstyle[12pt,aaspp4]{article}

\begin{document}

%definitions
\def \coto{$^{12}$CO(2--1) \/}
\def \cooz{$^{12}$CO(1--0) \/}
\def \thco{$^{13}$CO(1--0) \/}
\def \twco{$^{12}$CO \/}
\def \vlsr{$V_{LSR}$ \/}
\slugcomment{To appear in {\it The Astrophysical Journal}}

\title{The Episodic, Precessing Giant Molecular Outflow
from IRAS~04239+2436 (HH~300)}
\author{H\'ector G. Arce \& Alyssa A. Goodman}
\affil{Harvard--Smithsonian Center for Astrophysics, 60 Garden St.,
Cambridge, MA 02138}
\authoremail{harce@cfa.harvard.edu, agoodman@cfa.harvard.edu}

\begin{abstract}
We present the first set of detailed molecular line maps of the
region associated with the
giant Herbig-Haro flow HH~300, from the young star IRAS~04239+2436.
Our results indicate that the red lobe of the
HH~300 flow is depositing a fair amount of momentum 
($3.2~(\sin i)^{-1}$~M$_{\sun}$~km~s$^{-1}$) and kinetic energy
($2.6~(\sin i)^{-2} \times 10^{43}$~ergs)
over a notable volume ($\sim 11$\%) of its host dark cloud.
This makes HH~300 a key player in the evolution and fate of its parent cloud.
The redshifted molecular outflow
lobe  of HH~300 is 1.1~pc long and 0.3~pc wide, and has a very clumpy structure.
The density, velocity, and momentum distributions in the outflow 
indicate that the observed clumps arise from the prompt entrainment
of ambient gas.
Bow shock-shaped structures are associated with the clumps, and we hypothesize that the shocks are produced by
different mass ejection episodes. Lines drawn from IRAS~04239+2436 to each
of these clumps have different orientations on the
plane of the sky, 
and we conclude that HH~300 is a precessing and episodic outflow.

The observations include a map of the red lobe 
in the \coto line, with a 
beam size of 27\arcsec, and more extended maps of the outflow
region in the \cooz and \thco lines, with 45\arcsec \/ 
and 47\arcsec \/ beam sizes, respectively. 
Due to ``contamination''
by emission from another molecular cloud along the same light-of-sight,
we are not able to study the blueshifted lobe of HH~300.
The combined \cooz and \thco line observations
enable us to estimate the outflow mass 
accounting for the velocity-dependent opacity of the \cooz line.
This method is much more precise than using \twco data alone.
We obtain a steep power-law mass spectrum for HH~300, which
we believe is best explained by the evolution of the outflow mass kinematics.
In addition, our \thco observations
show that the HH~300 flow has been able to redistribute (in space and velocity)
considerable amounts of its surrounding 
medium-density ($\sim 10^3$~cm$^{-3}$) gas.

\end{abstract}

\section{Introduction}

A Herbig-Haro (HH) object is a nebulous knot which delineates the
shock arising from the interaction
of a high-velocity flow of gas ejected by a young stellar object (YSO)
and the ambient medium. A chain of these HH objects (or knots)
is usually referred as an HH flow.
Recently, it has been found that very long ---parsec-scale--- 
HH flows exist, 
and that they are quite common 
(Reipurth, Bally, \& Devine 1997, hereafter RBD).
This has sparked major interest on the impact of these energetic mass flows
on their environment, in particular the relation between optical
jets and molecular outflows and the physics of the entrainment mechanism.
Giant HH flows have sizes about an order of magnitude larger than 
the cloud cores from which they originate, and many are found to extend
to distances well outside the boundary of their parent dark cloud (RBD).
One or more shocks associated with the same
flow can accelerate the entrained gas to velocities greater than those
of the quiescent cloud, transferring momentum and energy into the molecular
cloud, and thus producing a molecular outflow.
The colossal size of a giant HH flow enables it to
entrain molecular material at parsec-scale
distances from the source, and thus it may affect the kinematics
and density of a substantial volume of its parent molecular cloud. 

Although several models try to explain the entrainment
mechanism that forms molecular outflows
(e.g., Shu et al.~1991; Masson \& Chernin~1993) 
and other numerous theoretical works investigate the effects YSO outflows 
have on their environment (e.g., Norman \& Silk 1981; 
Raga, Cant\'o, \& Steffen~1996;
Matzner \& McKee~2000),
there has yet to be a consensus reached on these issues.
On the observational front, recent high-resolution studies of molecular outflows
(e.g., Bence, Richer, \& Padman~1996; Cernicharo \& Reipurth 1996;
Lada \& Fich 1996; Davis et al. 1997; Yu, Billawala, \& Bally~1999,
Lee et al.~2000) have provided significant new constraints on the physical
parameters of outflows and the entrainment mechanism. 

In order to study
the effects an outflow has on its environment's kinetic and density structure,
it is imperative to observe a large area of the cloud gas 
surrounding the outflow, in addition to the outflow itself.
The observations should preferably be of more than 
one molecular line transition, probing a range of densities, 
and at least one of the lines should 
be relatively optically thin, in order to take line opacity into account.
In this paper, we present new $^{12}$CO(2--1), $^{12}$CO(1--0),
and \thco observations
of the HH~300 outflow, its surroundings and a large fraction of its parent 
dark cloud. 
 
The HH~300 flow was discovered, in the optical, 
by RBD.
HH~300 is located in the westernmost region of 
the B18 molecular cloud (usually called
B18w) in the Taurus cloud complex. 
The B18w region, also called IRAS core Tau G1 by
Wood, Myers, \& Daugherty~(1994), looks
like a 1.7 by 0.7~pc filamentary protuberance from the main cloud (B18),
see Figure~1.
The source of the HH~300 flow was established to be 
IRAS 04239+2436\footnote{Although this paper is the 
first one to present a map of the HH~300 molecular 
outflow, Moriarty-Schieven et al.~(1992) observed high-velocity wings
in the $^{12}$CO(3--2) line at the source position, which led them
to conclude that IRAS 04239+2436 is a molecular outflow source.} (RBD), which lies at the northern end of B18w, just
west of the ``bridge'' between B18 and B18w. 
Near- and mid-infrared photometry (Myers et al.~1987),
and its infrared spectrum (Greene \& Lada~1996)
indicate that this is a 1.3 L$_{\sun}$ Class I young stellar object.
The HH~300 flow was originally discovered to
consist of three redshifted HH objects (HH~300A, B, and C),
each with a bow-shock-like morphology, 
close together at a distance of about 1.1 pc southeast of the outflow source,
and a small blueshifted knot (HH~300D) about 0.02~pc
northwest from the source, assuming a distance to Taurus of 140~pc 
(Kenyon, Dobrzycka, \& Hartmann~1994). 
Recent NICMOS images at 1.644 and 2.122\micron \/, with a spatial resolution 
of $\sim 0.13\arcsec$,
of the HH~300 source region
by Reipurth et al.~(2000) reveal: a cometary nebula surrounding the source;
a jet on the blueshifted side of the HH flow (northwest of the source)
along the symmetry axis of the nebula; and that IRAS 04239+2436
is a binary.

\section{Observations}

\subsection{$^{12}$CO $J= 2 \rightarrow 1$ \/ Data}

The \coto line data were obtained using 
the on-the-fly mapping technique at the 
National Radio Astronomy Observatory (NRAO) 12~m 
telescope on Kitt Peak, Arizona, in December 1998. 
At the observed frequency of 230~GHz, the telescope's 
half-power beam width, main beam efficiency, and aperture efficiency 
are 27\arcsec \/, 0.32 and 0.44, respectively. 
The spectrometer used was a filter bank with
250 kHz resolution, with two independent sections of 128 channels each.
The filter bank was put in parallel configuration, in which each 
of the two sections received independent signals with a different
polarization. The parallel configuration was chosen so that ultimately
the two polarizations could be averaged to produce better signal-to-noise
spectra. At a frequency of 230~GHz, the resultant velocity resolution for this
setup is 0.65~km~s$^{-1}$.

The on-the-fly (OTF) mapping technique was used to observe an 
11\arcmin \/ by 37.5\arcmin \/ area, along a position angle of 49\arcdeg, surrounding
the axis of the HH 300 flow's redshifted lobe. The OTF technique 
allowed the extended area to be mapped in a more efficient way than the
conventional point-by-point mapping. The telescope in OTF mode
moved across the source at a constant speed of 30\arcsec/sec,
while data were taken a rate of 10 samples/sec.
In order to map the desired area more
efficiently, the total area mapped consists of
overlapping regions of different sizes: one large sized map 
of 8\arcmin $\times$ by 33\arcmin ,
one medium sized map of 8\arcmin \/ $\times$ 20\arcmin \/
and two small sized maps of 3\arcmin \/ $\times$ 6\arcmin \/ 
and 8\arcmin \/ $\times$ 5\arcmin .
All the regions were scanned along the direction perpendicular to the 
HH flow's axis, which has a P.A.$=49\arcdeg$. The separation, in the direction
perpendicular to the scanning direction, between subsequent rows was 6\arcsec. 
The telescope was pointed to an OFF position, located at R.A. $4^h21^m40^s$,
Dec. 24\arcdeg05\arcmin00\arcsec (B1950), after
every other row, where it would observe the OFF position for 10 sec and then
vane calibrate for 5 sec. The OFF position was later observed to have
emission in the velocity range between 2.4 and 6.3~km~s$^{-1}$,
with $0.1\lesssim T_{A}^{*} \lesssim 0.5$~K, 
and no emission greater than 0.1~K at other velocities. The
emission from the red lobe of the HH~300 outflow is at
velocities greater than 6.7~km~s$^{-1}$.
Therefore, the small amount of emission at velocities
between 2.4 and 6.3~km~s$^{-1}$ at the OFF position does not affect
our outflow data (see Figure~2).
The different regions were observed several times to improve the 
signal-to-noise in the spectra.  
The system temperature was measured to be in the range
between 400 and 900~K. 
In order to keep a constant noise level through the whole map,
the regions that were mapped with a higher system temperature were observed 
more than the regions with lower system temperature.
 
The raw OTF data were reduced using the various Astronomical
Image Processing Software (AIPS) OTF tasks.
A first-order baseline was fitted and subtracted to each spectrum,
and the two polarizations were averaged. The different mapped regions were 
combined and averaged. The map was then convolved onto a grid with
14\arcsec \/ pixels. The resultant rms noise 
in each 0.65~km~s$^{-1}$ channel was 0.12~K, for most (non border) spectra.
The intensity scale of the spectral
data presented in this paper is in units of $T_{A}^{*}$ (Kutner \& Ulich 1981) unless 
it is stated otherwise.

\subsection{$^{12}$CO $J= 1 \rightarrow 0$, $^{13}$CO $J= 1 \rightarrow 0$
\/ and C$^{18}$O $J= 1 \rightarrow 0$ \/ Data}

In order to study the effect of the HH~300 outflow on a larger scale and to attain
a better estimate of the outflow mass, we observed
the \cooz and \thco lines in a region 
17.5\arcmin \/ by 59\arcmin \/ along a position 
angle of 49\arcdeg, surrounding
the HH~300 outflow source (IRAS 04239+2436).
The data were obtained using the SEQUOIA 16 element 
focal plane array receiver of the Five College Radio Astronomy Observatory
(FCRAO) 14~m telescope. The observations were done over the course of three
different observing rounds, which took place in April 1999, December 1999,
and February 2000. Both lines were observed in frequency-switched mode and
the backend used was the Focal Plane Array 
Autocorrelator Spectrometer (FAAS), with a channel spacing of 78~kHz 
(0.21~km~s$^{-1}$) for $^{12}$CO(1--0), at a frequency of 115~GHz,
and 20 kHz (0.05~km~s$^{-1}$) for $^{13}$CO(1--0), at a frequency of 110~GHz.
The telescope half-power beam width for the \cooz and \thco lines
are 45\arcsec \/ and 47\arcsec, and the main beam efficiencies are
0.45 and 0.54, respectively. The \cooz spectra were taken with a spacing of
44\arcsec \/ (beam sampled) and with an integration time of 150~sec for
each position. The \thco spectra were taken with a spacing
of 22\arcsec \/ (Nyquist sampled), and with an integration time of 100~sec
for each position.
The system temperature of our observations 
ranged between 700 and 1000~K for $^{12}$CO(1--0),
and between 300 and 500~K for $^{13}$CO(1--0). 

Observations of the C$^{18}$O(1--0) emission were also made using the same 
telescope and backend configuration as the $^{13}$CO(1--0) observations.
We made three Nyquist sampled 
FCRAO-SEQUOIA footprints. The 5.5\arcmin \/ $\times$ 5.5\arcmin \/ 
footprints were centered at offsets (0,0), (-3.1\arcmin,-3.1\arcmin), and
(-6.9\arcmin, -9.4\arcmin) with respect to the outflow source position at
($\alpha$,$\delta$)$_{1950}$ = ($4^h23^m54^s.5$, 24\arcdeg36\arcmin54\arcsec).
The telescope half-power beam width and main beam efficiency for 
C$^{18}$O(1--0), at a frequency of 109~GHz, are 47\arcsec and 0.54,
respectively. Each position was observed for 200 to 300 sec, with a system
temperature ranging from 170 ~K to 350~K.

The \cooz and \thco
data were reduced with the CLASS and MIRIAD packages. The original 
spectra from both lines were smoothed and resampled to a
channel spacing of 0.22~km~s$^{-1}$. The spectra were spatially linearly 
interpolated and a data cube of 11.5\arcsec \/
pixels was produced for each line. The data cube was then smoothed 
by convolving it with a gaussian with a FWHM of 46\arcsec, using the
task {\it smooth} in MIRIAD. The resultant rms noise of the spectra in
the \cooz and \thco smoothed maps
was 0.17~K and 0.07~K for each 0.22~km~s$^{-1}$ channel, respectively.
Interpolating the \cooz and the \thco data to this common position-velocity
grid allowed us to use them in concert when calculating the mass estimates 
presented in \S 3.3.2.

\section{Results}

Figure~2 shows spectra of  $^{12}$CO(2--1), 
$^{12}$CO(1--0), and \thco averaged over the area mapped in $^{12}$CO(2--1). 
The most notable property of the spectra in Figure~2 is that they are
composed of two main (gaussian-like) components. 
One component is centered at a velocity close
to 6.7~km~s$^{-1}$ and the other component is centered at a velocity close
to 5~km~s$^{-1}$. The $v=6.7$~km~s$^{-1}$ component is produced by the molecular emission
of the HH~300 outflow host cloud (B18w), while the other component,
at $v=5$~km~s$^{-1}$,
is produced by another cloud in Taurus (hereafter cloud A), along the same 
line-of sight.  Unlike many other average outflow spectra 
(e.g., HH~111, Cernicharo \& Reipurth (1996);
NGC~2264G, Lada \& Fich (1996); RNO~43, Bence et al. 1996),
the red wing of the \coto and \cooz average spectra 
is not very prominent. The wing only extends from about 8 to about 10~km~s$^{-1}$.
In addition to the small red wing, the line core component of the B18w cloud, 
seems to have a slight asymmetry, with more emission towards red velocities. 
This suggests that the HH~300 outflow is composed mainly of gas with
slow radial velocities, and the gas emission of its slowest velocities might
be hidden under the ambient cloud emission. 

The cloud A component is visible not only in the area surrounding 
the red lobe of HH~300, but through most of the area we mapped in 
\cooz and $^{13}$CO(1--0). This area includes the region 
where the blue lobe of HH~300
should be, assuming that it lies northeast of IRAS~04239+2436,
opposite to the red lobe, at the same position angle.
There is evidence in the optical (RBD) and near-infrared
(Reipurth et al.~2000) for 
blueshifted outflowing material only very close to the source.
But, even if the HH~300 flow's blue lobe were interacting with the surrounding
gas (close to and/or far from the source), we would not be able to study it.
The ``contamination'' from cloud A on the blue side of 
the B18w component is such that we would not be able to observe the 
high-velocity blue wings that presumably would have been present in the
\twco spectra. Therefore we concentrate our study on the
interaction between the redshifted lobe of the HH~300 flow and its surroundings.

\subsection{$^{13}$CO $J= 1 \rightarrow 0$ Maps}

Figure~3 shows five \thco velocity maps, integrated over
five different velocity ranges. The first map is integrated over the velocity
in which the emission from cloud A 
is present, $3.79<v<5.77$~km~s$^{-1}$. The other three maps are made over velocity 
ranges in which there is emission from the B18w cloud (and HH~300 outflow).
It is evident that the velocity map in Figure~3a 
shows a \thco structure different from the other velocity maps.
In addition, the position angle of the short and long axes
of the  $3.79<v<5.77$~km~s$^{-1}$ structure differs from that of the
B18w cloud by about 30\arcdeg.
This reconfirms our assumption that the emission from velocities 
3.8 to 5.8~km~s$^{-1}$ does not come from the B18w cloud, but from
another cloud, cloud ``A'', on the same line-of-sight. 

The integrated 
velocity map over the range $5.77<v<7.09$~km~s$^{-1}$ is shown in Figure~3b.
The velocity range of Figure~3b does not cover the whole range
of velocities in which there is emission from the B18w cloud component
($5.77<v<8.63$~km~s$^{-1}$) in Figure~2.
But, even though the map in Figure~3b only shows emission
from 5.77 to 7.09~km~s$^{-1}$, it
shows the general \thco structure of the B18w cloud. 
Comparing Figure~3b with an integrated map which covers the whole
velocity range of emission from the B18w component ($5.77<v<8.63$~km~s$^{-1}$)
shows that both maps have essentially the same
structure. And, comparing Figure~3b with a map 
integrated over just 6.2 to 7.1~km~s$^{-1}$ also
shows that they have similar structure. In addition, the shape of these \thco
maps is very similar to the shape of the B18w dark cloud as seen in the optical
image of RBD (see Figure~6). 
Therefore, the emission from the B18w  
cloud is predominantly from gas
at velocities between 6.2 and 7.1~km~s$^{-1}$, the peak emission range for the 
spectra in Figure~2.
 
The integrated intensity map over $7.09<v<7.53$~km~s$^{-1}$ 
has a very peculiar structure (Figure~3c). The long
and short axis, as well as the position angle of this structure are similar
to those of the $5.77<v<7.09$~km~s$^{-1}$
(Figure~3b) structure. Also, the $7.09<v<7.53$~km~s$^{-1}$
structure bears some resemblance to the \cooz 
outflow in the velocity range of $7.53<v<7.97$~km~s$^{-1}$ 
(see Figure~5a).
The local maximum in \thco at R.A. $4^h22^m50^s$, 
Dec. 24\arcdeg21\arcmin30\arcsec (B1950)
is at the tip of the most prominent bow shock-like clump (clump R3, see
section 3.2, and Figures~3, 4 and 5)  in the $^{12}$CO
outflow. Although it is tempting
to conclude that most of the emission at these velocities comes
from the ``slow-moving'' outflowing gas, the velocity of this structure
(about 7.75~km~s$^{-1}$) is still very close
to the line center velocity ($V_{LSR}=6.67$~km~s$^{-1}$). Thus, we expect
some contribution to the emission at $7.09<v<7.53$~km~s$^{-1}$ to come from
the turbulent gas in B18w unaffected by the outflow. 
We are convinced that \thco emission from both slow-moving
outflowing gas and ambient cloud gas 
are present in the integrated intensity 
map in the velocity range of $7.09<v<7.53$~km~s$^{-1}$ 
(see \S 3.3.3 for a quantitative discussion). 

Figure~3d shows the integrated intensity for the velocity range 
$7.53<v<7.75$~km~s$^{-1}$. There is no doubt that the \thco emission in this
velocity range, southwest of IRAS~04239+2436 is due to gas that has been put 
in motion by the outflow. The \thco emission coincides in position 
with the medium-velocity structure seen in the \twco outflow velocity maps 
(see Figures~3 and 4). The emission detected northeast of the outflow source at
these redshifted velocities, with respect to the ambient cloud velocity
($V_{LSR}=6.67$~km~s$^{-1}$),
is not associated with the HH~300 flow. Instead, this emission comes from an
elongated structure perpendicular to the HH~300 flow's axis, associated
with the B18 molecular 
cloud, as can be seen in the large scale, 2.7\arcmin \/ resolution,
\thco velocity maps of Taurus by Mizuno et al.~(1995).  The integrated
intensity map over the range $7.75<v<8.63$~km~s$^{-1}$ (see Figure~3e)
shows the highest (red)
velocities of the HH~300 outflow in \thco emission. The knotty features of these
high velocity \thco coincide with the position of the high-velocity features
in the \twco outflow (see Figures 4 and 5). These knotty features are well defined in 
space and velocity.  

\subsection{$^{12}$CO Maps}

In Figures 4 and 5 we show \coto and \cooz velocity
maps of the HH~300 molecular outflow.
The \cooz data set covers more area and has a better spectral resolution
than the \coto data. 
On the other hand, the \coto data has a better spatial resolution
and sensitivity than the \cooz data. 
Regardless of the differences in spectral and spatial resolution, area and
sensitivity, the \coto and \cooz maps 
look very similar for similar velocity ranges (see Figures 4 and 5).
The most notable characteristic of this outflow is its very clumpy
structure. We identify five molecular
clumps in the redshifted lobe of the HH~300 outflow.
In Figure~6 and Table~1 we show how five named clumps (R1 to R5) are identified.

We believe that all five clumps mentioned above are associated with 
the HH~300 outflow and each clump represents a different ejection
event. Clumps R1 and R2 have ``blobby'' 
structures like the high-velocity CO bullets found in HH~111
by Cernicharo \& Reipurth (1996). 
Clump R3 has a shape reminiscent of a bow shock, with wings pointing
in the direction of the outflow source (see Figures 4b and 5b).
There is stronger \twco emission in the south wing
than in the north wing of the bow structure. This asymmetry in the bow 
wings of R3 is most probably due to the underlying distribution of molecular
gas in the cloud, since there is more gas emission south of clump R3,
than to its north. 
Recent high spatial resolution observations of outflows 
(e.g., Yu et al.~1999, Lee et al.~2000) have found that  
bow shape CO structures are common. The CO bow structures presumably are
created by the interaction between a bow shock from an underlying jet, and
the ambient gas.
Clump R4 is barely distinguishable
as a separate clump from R3, but 
it is still clear that R4 exhibits a local maximum
of CO emission and that it diverges from the bow structure of R3 
(see Figure~4b). 
We are confident
that R4 is associated with the HH~300 outflow not only due to its
relatively (redshifted) high-velocity, but also because it coincides in 
position with the optical knot HH~300C (see Figure~6).
The molecular clump R5 is the clump with the slowest
velocities, and can only be seen in the velocity maps with velocities close
to the cloud velocity. It has an elongated drop-like structure, as if it 
were a bow-shock, with only one wing. The same effect that produces
asymmetric bow wings in R3, could be producing the structure of R5, but in a
more dramatic way. In the case of R5, there is practically no molecular
emission north of the clump, but there is some emission south of it. It is
very probable that we do not see a north bow wing in R5, since there
is essentially no gas to the north of the clump to produce a \twco wing.
Table~2 shows the
distance from the source and the position angle, as well as other
physical properties of each of the molecular clumps.
  
\subsection{Outflow Physical Properties}

In order to study the effects of a molecular outflow  on its environment,
it is essential to calculate its mass, momentum, and energy in the most 
precise way possible. There are several potential pitfalls which,
if not avoided or compensated for, can reduce the precision of
the calculation of an outflow's physical properties. Major dangers include:
(1) not accounting for the inclination of the flow's axis to 
the plane of the sky; (2) assuming that the molecular spectral
line being used to obtain the
physical properties is optically thin when in reality 
it is optically thick; and (3) not counting slow-moving outflow gas which is
``hidden'' by the ambient cloud molecular emission as part of the flow.
With our data, we have found
a way to tackle the uncertainties produced by
(2) and (3). On the other hand, (1) remains a problem.

\subsubsection{Inclination} 
 
The most accurate way of obtaining $i$ for a molecular outflow is to observe 
the tangential ($V_t$) and radial velocities ($V_r$)
of an HH object associated with the molecular outflow being studied.
Assuming that the HH object travels through space along a straight line and
that the molecular gas from the molecular outflow 
associated with the HH object
will move in the same direction, then 
$i=\arctan (V_t/V_r)$.
 Unfortunately, no proper motions study of the HH~300 
optical knots has been done, thus we do not know the transverse
velocities of the HH~300 knots. 
Other methods of estimating the value of 
$i$ for molecular outflows use a
simple geometrical and velocity field model for the flow,
combined with the observed data of the outflow 
(see Liseau \& Sandell~1986; Cabrit et al.~1988). 
Using these models results in $i$ values
with large uncertainties, but 
at least they give a rough estimate to the value of $i$, when otherwise
there would be none. The clumpy structure of the
HH~300 molecular outflow
(very different from the molecular outflow cone geometry 
assumed by the models listed above), and the fact that it is impossible to 
detect any blueshifted CO emission from the HH~300 outflow due to contamination
from cloud A, makes it impossible to use the simple geometric
models to estimate 
$i$ for HH~300. In addition, if the HH~300 outflow is precessing and episodic 
(see \S 4.3),
then each mass eruption could have been ejected 
at a different angle with respect 
to the plane of the sky, so that using a single value 
of $i$ for HH~300 would be an incorrect assumption. 

The HH~300 flow most likely lies very close to the plane of the sky.
Most of the giant HH flows for which the inclination angle is known
have been found to have small values of $i$.
By definition, giant HH flows are flows which are observed to
extend more than 1~pc in length on the sky. The smaller the angle between
the flow's axis and the plane of the sky, the longer the projected
(on the plane of the sky) size of a flow will be, for a given flow length. 
Thus, it is not surprising that there is a bias toward small values of $i$ 
in giant HH flows. The small maximum outflow radial
velocities ($\sim$ 3 to 4~km~s$^{-1}$)
that the HH~300 molecular outflow exhibits,   
as well as the morphology of the near-infrared nebula surrounding the outflow 
source (as observed by Reipurth et al.~2000) further confirm
that HH~300 lies very close to the plane of the sky. We believe that
the (average) axis of the HH~300 flow must have an inclination angle 
with respect to the plane of the sky between 5 and 15 degrees. In this paper,
all outflow velocities quoted for HH~300
{\it are not corrected by inclination angle}, unless otherwise noted.

\subsubsection{Opacity Correction}

Using an 
optically thick line to study an outflow will cause an underestimation  
of the mass and kinetic energy in the flow.
Not only should one worry about the line core 
opacity, but also the dependence of the line opacity on velocity.
Using the outflow mass estimation methods of 
Bally et al.~1999 (hereafter BRLB) and
Yu et al.~1999 (hereafter YBB), 
with our \thco and \cooz data in HH~300, 
we believe we have obtained a fairly
accurate estimate of the outflow mass. 

We do not strictly follow BRLB's or YBB's method, but use a combination of the
two, with our own modifications. First, we take an average of all the 
\thco and \cooz spectra
in the area southwest of IRAS 04239+2436,
where the outflow lies. Then, we plot the ratio
of the lines as a function of velocity (see Figure~7a).
The \cooz line is optically thick over all velocities where there is detectable
$^{13}$CO(1--0), as the line ratio is always less than 60.
It is clear that the 
line ratio (and thus the opacity) is not constant over velocity, 
and that it has a parabolic shape. Assuming a constant opacity
for the \cooz line,
as in most studies which claim to correct for the \cooz opacity
using the \thco line,
is {\it not} an accurate assumption. It is imperative to
take the velocity dependence of the opacity
into consideration.

Using the average spectra of the two molecular lines,
we obtain the value of the main beam temperature ($T_{mb}$)
for each  0.22~km~s$^{-1}$ wide channel,
and calculate the ratio of main beam temperatures. 
We then make a second-order polynomial fit to the ratio of \cooz to \thco
intensity, $R_{12/13}$, as a function of velocity.
The fit is constrained to have a minimum
at the velocity of the ambient cloud (equal to the
velocity where the average \thco spectrum peaks), \vlsr = 6.67~km~s$^{-1}$. 
We exclude the five velocity channels closest to the ambient velocity
from the fit, 
as they are the velocities at which the $^{13}$CO emission is slightly
optically thick (see below)
and the $^{12}$CO emission is extremely optically thick.
Also, we use only the points that lie between the velocities 5.8 and 8.2~km~s$^{-1}$,
in order to exclude ``cloud A'' and low signal-to-noise velocities.
We use the $R_{12/13}(v)$ parabolic fit to extrapolate 
$R_{12/13}$ to the high-velocity wings
of the outflow, where the \thco line is too weak to be reliably detected.
 
We estimate the mass in the following way.
If the \thco emission at a velocity $v_i$, at a certain
position ($x_i$, $y_i$) is greater than or equal to 
twice the rms noise of the spectrum at that position, 
we use the main beam temperature value at the given velocity 
($T_{mb}^{13}(x_i, y_i, v_i)$).
If the \thco emission is less than 
twice the rms noise, then we use the \cooz spectrum at the given position 
($x_i$, $y_i$) and velocity $v_i$ to estimate $T_{mb}^{13}(x_i, y_i, v_i)=T_{mb}^{12}(x_i, y_i, v_i)[R_{12/13}(v_i)]^{-1}$.
Similar to what BRLB and YBB do, the function $R_{12/13}(v)$ is truncated
at a value of 62, the assumed isotopic ratio (Langer \& Penzias 1993). 
If both the \thco and the \cooz emissions at ($x_i$, $y_i$, $v_i$)
are less than 
twice the rms noise, then essentially there is no signal from which we can
estimate a value of  $T_{mb}^{13}(x_i, y_i, v_i)$.

Once we obtain the value of $T_{mb}^{13}(x_i, y_i, v_i)$,
 we can get the value of the opacity, assuming
that the \thco line
is optically thin. Using our C$^{18}$O data we are able to estimate the opacity
of the \thco emission at several positions. The C$^{18}$O
observations were made 
where strong \thco emission was found; (a.k.a.) regions where the 
\thco emission might be optically thick.
From the $^{13}$CO to C$^{18}$O ratio
we estimate the opacity of the \thco line (as a function of velocity) at the 
limited positions for which we have C$^{18}$O data. The $^{13}$CO to
C$^{18}$O ratios indicate that 
even in regions with strong $^{13}$CO emission, the opacity of the \thco
line at line core velocities ($6.2 < v < 7.0$ km~s$^{-1}$)
does not exceed a value of 3, and for other 
velocities the $^{13}$CO emission is optically thin. Since we are mainly 
interested in the outflow mass our assumption
that \thco is optically thin is an excellent assumption.

At each position and at each velocity for which we have an estimate
of $T_{mb}^{13}(x_i, y_i, v_i)$, we calculate the \thco
opacity, $\tau_{13}(x,y,v)$, using: 
\begin{equation}
\tau_{13}(x,y,v) = -{\rm ln}[1 - \frac{T_{mb}^{13}(x,y,v)}{\frac{T_o}{\exp(T_o/T_{ex})-1} - 0.87}]
\end{equation}
from (Rohlfs \& Wilson~1996). 
Here $T_o = h\nu/k$, which for \thco is 5.29. 
The value of $\tau_{13}$, from equation~1,
agrees with the value of the $^{13}$CO
opacity obtained using the $^{13}$CO to C$^{18}$O line ratio in the 
limited number of positions where we measured the C$^{18}$O emission.
The excitation temperature, $T_{ex}$,
is obtained by assuming that the  \cooz line core is optically 
thick. We take an average spectrum of the \cooz data
over all of the area southwest of the source for which we have \cooz data,
and measured the peak temperature (corrected for main beam efficiency) 
to be $T_{peak}=9.6$~K. Solving the radiative transfer equation for the
excitation temperature, assuming that the line is optically
thick, we obtain this expression:
\begin{equation}
T_{ex} = \frac{5.53}{{\rm ln\/}[1 +\frac{5.53}{T_{peak}+0.82}]}
\end{equation}
from (Estalella \& Anglada 1997). The resultant
$T_{ex}$ is 13~K, the same value that Mizuno et al.~(1995)
quote for the whole Taurus molecular cloud complex.
Then, at each position and velocity the 
column density, $N_{13}(x,y,v)$, 
can be calculated with the following equation 
from Bourke et al.~(1997):  
\begin{equation}
N_{13}(x,y,v)=2.42\times10^{14}(T_{ex}+0.88)\frac{\tau_{13}(x,y,v) dv}{1-\exp(-T_o/T_{ex})}
 \end{equation}
The outflow (molecular hydrogen) mass at each position pixel and
 each velocity channel is given by 
\begin{equation}
M(x,y,v)= m_{H_{2}} N_{H_{2}}(x,y,v) A
\end{equation}
where $m_{H_{2}}$ (which is equal to 2.72 times the mass of a hydrogen atom)
is the mean molecular weight of the gas, $N_{H_{2}}$ is the molecular 
hydrogen column density which is obtained using the relation
$N_{H_{2}}=7\times10^5 N_{13}$ (Frerking, Langer, \& Wilson ~1982), 
and $A$
is the physical area of the pixel at the
distance of the source. We then sum over the area of the outflow
($M(v)=\sum_{area} M(x,y,v)$)
 to obtain the outflow mass as a function of velocity, and then sum
over velocity ($M_{total}=\sum_{v} M(v)$) to obtain the total outflow mass. 
The uncertainty in the mass estimates is limited to the uncertainty in the
$N_{H_{2}}$ to $N_{13}$ ratio, thus the uncertainty in the mass estimates
is about a factor of 2 (Frerking et al.~1982).

Using this method we can calculate
the mass per 0.22~km~s$^{-1}$-wide
velocity channel ($dM(v)/dv$) in the HH~300 flow.
In Figure~7b 
we plot our results for outflow velocities ($v_{out}$) greater than
0.5 km~s$^{-1}$, where $v_{out}$ is the observed velocity ($v$)
minus the cloud velocity
($V_{LSR}=6.67$).
Figure~7b shows that the  
observed mass (per 0.22~km~s$^{-1}$-wide velocity channel), corrected only for the
velocity-dependent opacity of the \cooz line,
has a power-law dependence on velocity for $v_{out}$ between
0.75 and 1.85~km~s$^{-1}$. For outflow velocities higher than 
1.85~km~s$^{-1}$, the mass has an even steeper power-law dependence. A power-law
fit to each of these two trends yields a slope of $-5.2\pm0.1$ for 
the low-velocity points,
and a slope of $-7.8\pm0.4$ for the high-velocity points.

\subsubsection{Correction for Ambient Cloud Emission}

We suspect that some  
velocity channels
have emission from both the low-velocity outflowing gas and the ambient 
cloud. Thus, in order to obtain an accurate estimate of the outflow mass it is
essential to correct for the ``extra mass'' at low  velocities
due to ambient cloud ``contamination''.  
In order to make this correction, we use
our calculations of the mass per velocity channel (\S 3.3.2) for 
$4.5<v<9.8$~km~s$^{-1}$. Using the overall (cloud A + B18w + HH~300) 
 mass spectrum in Figure~7c, we fit a two-gaussian function to
the points which define the clouds (cloud A and B18w) and we are sure 
are not ``contaminated'' by the HH~300 outflow (see Figure~7c). 
The gaussians, which seem a good 
fit to the relevant ambient emission, give us a value of 
$V_{LSR}=6.67$~km~s$^{-1}$
as the center velocity of the B18w component, which is the value
we use as the velocity of the cloud throughout the paper, and a 
center velocity for cloud A of $4.95$~km~s$^{-1}$.
The outflow emission can clearly be seen in Figure~7c 
as a deviation from the gaussian shape at
redshifted velocities. To obtain the corrected
outflow mass ($M_{cor}(v)$), we subtract the value
of the fit at the velocity of interest ($M_{fit}(v)$) from the observed mass 
($M_{obs}(v)$). Using the simple relation $M_{cor}(v)=M_{obs}(v)-M_{fit}(v)$ we
obtain an outflow mass which is presumably free from the cloud emission.
We display the mass-velocity relation using the corrected mass in Figure~7d.
It can be seen that the low-velocity power-law slope of the
mass spectrum (Figure~7d) becomes shallower,
with a slope of $-4.0\pm0.2$, after the correction.
From here on all the outflow mass (and momentum) values given are corrected
for ambient cloud emission, unless otherwise noted.

We note that this method will result in an underestimation of mass in the
lowest outflow velocity channels, if the emission were 
extremely optically thick at these velocities
or if the points used for the ``cloud'' fit were 
heavily affected by outflow emission. As discussed above, neither of these two
scenarios is our case, so we are confident that our underestimation  
of low-velocity outflow mass is minimal.
The mass and the power-law slope at high outflow velocities 
($v_{out} > 1.85$~km~s$^{-1}$)
are unchanged after the ambient cloud correction, since 
there is no contamination from the ambient cloud at those velocities.

\subsubsection{Effects of Inclination, Opacity and Ambient Contamination}

In estimating an outflow's momentum and energy, corrections for inclination 
can have profound effects, since we can only {\it observe} a radial velocity. 
On the other hand, the slope of the mass spectrum (see Figure~7b)
is not affected by a flow's inclination. Correcting the velocity for 
projection effects will
just re-scale the velocity axis of the mass spectrum
plot, but the slope of the power-law mass spectrum will remain the same, 
assuming that all parts of the flow have roughly the same inclination angle,
$i$. 

The effects that  velocity-dependent opacity correction have on the outflow's
mass and mass spectrum can be seen clearly in Figure~7b. In this figure we
plot the mass spectrum of the red lobe of HH~300, 
using three different assumptions. The square symbols denote the mass spectrum
obtained assuming that the \cooz line is optically thin, so that
mass is proportional to $\int T_{mb} dv$. The diamond symbols represent 
the mass spectrum obtained by assuming that the \cooz line opacity, and thus
$R_{12/13}$= $I$($^{12}$CO)/$I$($^{13}$CO), is constant.
We use the area- and velocity-averaged value of $R_{12/13}$, over
the velocity range between 7.3 and 8.5~km~s$^{-1}$ (or outflow velocities
between 0.75 and 1.85~km~s$^{-1}$) as an average opacity, $\tau_{avg}$.
The mass is then calculated using equations 1, 3, and 4, and assuming
that $T_{mb}^{13}=T_{mb}^{12}/\tau_{avg}$, with $\tau_{avg}=13$.
 The triangle symbols show the mass spectrum
obtained using the technique described in section 3.3.2. 

From Figure~7b it 
is clear that assuming \cooz is optically thin greatly underestimates the
mass (by a factor of ten) compared to the opacity-corrected estimates. 
The constant opacity correction gives a more realistic estimate of the
total mass. The shape and slopes of the 
mass spectrum using the constant opacity correction 
and the mass spectrum obtained using the optically thin assumption
are basically the same.
 Assuming that
the opacity of the \cooz is constant implies that the high-velocity
\cooz emission is optically thick. 
Since it is clear that the \cooz line is optically
{\it thick} at outflow velocities
close to the cloud $V_{LSR}$ and it is optically {\it thin}
at high velocities, a 
more precise approach is to assume that the \cooz line opacity depends
on velocity. The velocity-dependent opacity correction results in
low-velocity mass estimates which are close to the constant opacity 
assumption, and at high velocities the mass estimates converge to the
optically thin assumption estimates. This steepens the mass spectrum 
in both the low and high velocity regimes.

Comparing Figures~7b and 7d we see that the correction for ambient cloud
emission makes the mass spectrum shallower.
This is because there is more contamination from the ambient
cloud at lower velocities than at higher velocities. Not correcting for such
contamination results in an overestimation of the mass at lower velocities. Thus, correcting for a 
velocity-dependent opacity and correcting for ambient cloud emission have
opposite effects on the mass spectrum slopes. Most previous outflow studies
do neither of the two corrections, so it could be that their estimates of the
mass spectrum slope are reasonable by chance.

\section{Discussion}

\subsection{Mass and Energetics}
The total mass of the red lobe of the HH~300 outflow, using the method 
described in sections 3.3.2 and 3.3.3,
for $v_{out} \ge 0.75$~km~s$^{-1}$, is 2.4~M$_{\sun}$. 
As can be seen in Figure~7d, 
the mass-velocity relation at velocities between 0.75 
and 1.85~km~s$^{-1}$ follows a power-law
trend, $dM(v)/dv ~ \alpha ~ v^{-\gamma}$. 
A second power-law, with a steeper slope, is present for velocities
greater than 1.85~km~s$^{-1}$. This broken power-law 
trend is typical of many outflows
(see \S 4.2).
If we take the ``outflow mass''  to include only  mass 
at velocities which follow the broken power-law trend, then we find
2.4~M$_{\sun}$ in the outflow. Notice the lone point at $v_{out}=0.53$~km~s$^{-1}$,
which lies below the extrapolation of the power-law trend in Figure~7d.
If we add the mass at this very slow outflow velocity 
(corrected by ambient emission, see \S 3.3.3), we then find  
the total mass of the red lobe of HH~300 outflow to be 4.3~M$_{\sun}$, for 
$v_{out} \geq 0.53$~km~s$^{-1}$. The mass of the ambient cloud 
gas that surrounds the red lobe of HH~300 (only the B18w gas southwest of
IRAS~04239+2436) is 57~M$_{\sun}$, thus the mass in the outflow, including
the slowest velocity material we can reliably call ``outflow'', is 
7\% of the mass of its surrounding cloud.

The total (line-of-sight) 
momentum in the red lobe of the HH~300 molecular outflow,
for $v_{out} \geq 0.75$~km~s$^{-1}$ is 2.2~M$_{\sun}$ km~s$^{-1}$, and for
$v_{out} \geq 0.53$~km~s$^{-1}$ is 3.2~M$_{\sun}$~km~s$^{-1}$. The amount of momentum
that has been deposited in the cloud by HH~300  presumably should be
substantially more, as we have only considered 
the line-of-sight velocity component. If we assume that $i$ is
between 5\arcdeg \/ and 15\arcdeg, then the outflow momentum would
be between 12 and 37~M$_{\sun}$ km~s$^{-1}$, for 
$v_{out} \geq 0.53$~km~s$^{-1}$. Thus, our ignorance of
$i$ brings huge uncertainties to the value
of the outflow momentum. 
The kinetic energy of the outflow is even more uncertain, as it depends
on $(\sin i)^{-2}$. The
kinetic energy of the outflow
for $v_{out} \geq 0.75$~km~s$^{-1}$ is 
$2.1~(\sin i)^{-2} \times 10^{43}$ ergs,
and for $v_{out} \geq 0.53$~km~s$^{-1}$ is 
$2.6~(\sin i)^{-2} \times 10^{43}$ ergs. So, for example, if $i=10\arcdeg$,
the kinetic energy is about $8 \times 10^{44}$ ergs.

Independent of the real (angle-corrected)
momentum, we can still map  line-of-sight
outflow momentum. In Figure~8
we show a contour plot of the total 
momentum for $v_{out}\geq 0.97$~km~s$^{-1}$ ({\it not}
corrected for inclination angle).
 We are sure there
is very little or no emission from the ambient cloud at these velocities.
 The momentum plot shows distinctive 
peaks along the outflow axis. These peaks coincide
with the different clumps discussed in section 3.2. The peaks are surrounded 
by a $\sim 0.3$~pc wide and $\sim 1$~pc long, lower-level, momentum distribution.

We assume that the dimension
along the line of sight of both B18w and the HH~300 outflow 
are similar to their respective short axes.
Using the maps in Figure 8,
we estimate their volumes to be $1.7\times0.7\times0.7 \sim 0.83$~pc$^{3}$
and $1.0\times0.3\times0.3 \sim 0.09$~pc$^{3}$, respectively.
Hence, we may
conclude that a notable fraction (0.09~pc$^3$/0.83~pc$^3$ $\sim$ 
11\% of the volume) of the 
B18w filamentary dark cloud is being injected with momentum from the
red lobe of the HH~300 outflow.  The mass, momentum, and kinetic energy
estimates of the HH~300 outflow are listed in Table~3.

\subsection{Mass-Velocity Relation}
An interesting aspect of the HH~300 outflow is the steep power-law slopes 
we find in the mass-velocity relation ($\gamma=4.0$ for $v_{out}$ 0.75 to
1.85~km~s$^{-1}$, and $\gamma=7.8$ for $v_{out} \geq 1.85$).
All molecular outflows exhibit a power-law mass-velocity relation in which
there is more outflowing mass at low velocities than at higher velocities,
that is, $dM(v)/dv ~ \alpha ~ v^{-\gamma}$, where $\gamma>0$.
Most outflows show a break in the power-law relation and a steeper
power-law dependence at high velocities.
Recently, Richer et al.~(2000)
compiled existing data on 22 molecular outflows 
with central source luminosity ranging from 0.58 to 
$3 \times 10^5$~L$_{\sun}$.
The low-velocity ($v<10$~km~s$^{-1}$) mass
spectra $\gamma$'s for the 22 sources ranges from 
about 0.25 to 2.5, with most of the values concentrating
between 1.25 and 2. For high velocities, 
$\gamma$ ranges from 2.5 to 7.5, with most of the values between 3 and 4. 

Numerical and theoretical studies have also shown that the mass spectrum
has a power-law dependence. The models of Zhang \& Zheng (1997) and 
Smith, Suttner, \& Yorke~(1997), in which a
 molecular outflow is produced by the 
entrainment of the ambient gas by a bow shock,  find
a $\gamma$ of about 1.8, and
steepening of the mass-velocity relation at high velocities. 
Downes \& Ray (1999), with their numerical bow shock models, find that
the slope of the power-law mass spectrum 
ranges between $\gamma=1.58$ and $\gamma=3.75$
(depending on the parameters of the simulation).
 Both Smith et al.~(1997) and 
Downes \& Ray~(1999) find that $\gamma$ increases with time.
On the other hand, the analytical
study by Matzner \& McKee~(1999) finds that $\gamma = 2$ for all times, 
independent of the time history of the driving wind and 
for a wide variety of ambient
density distributions. The study of Matzner \&  McKee~(1999)
differs from the other ones listed above in that 
the outflow is not specifically created by the entrainment
of the ambient gas by a bow shock, but rather by a wide-angle 
wind which sweeps the ambient gas into a momentum-conserving shell.

The majority of observations and models in the literature to date
seem to indicate that $\gamma \sim$ 2.  Our observations
indicate that the HH~300 outflow has a much steeper ($\gamma \sim 4$) 
mass spectrum than most outflows (from low and intermediate-mass stars),
both observed and modeled. On the other hand,
our results are similar to the outflows analyzed by BRLB, YBB, and 
Yu et al.~2000 (hereafter, YBSBB). 
In Table~4 we show the average value of $\gamma$ for 
the outflows where mass estimates have included a correction
for velocity-dependent $^{12}$CO line opacity.  
From Table~4 we can see that, after this correction, 
most of the outflows have power-law
slopes larger than 2.
A small difference between the results in this paper and those of 
BRLB and YBB is that we find a steepening of the mass spectrum power-law
for high velocities in HH~300, whereas BRLB and YBB state that 
the outflows in their studies do not show such steepening.
Examining the mass spectrum plots of BRLB and
YBB, it is clear that some of them do not show a break in the power-law. Yet,
others of their 
mass spectrum plots could be fitted with two 
distinct power-laws, where the high-velocities would have a 
steeper $\gamma$ than the low-velocities.

There is no question
that  the \twco line is optically thick in the HH~300 flow,
and that its opacity depends 
on velocity, as the $^{12}$CO(1--0)/$^{13}$CO(1--0)
plot (Figure~7a) demonstrates.
The same is true for
the Circinus (BRLB), the B5-IRS~1 (YBB) and the MMS~9 (YBSBB)
outflows.
It is safe to say that the \twco line
of most, if not all, outflows is optically thick at low velocities and that its 
opacity is velocity dependent. Thus, if low-velocity 
$^{12}$CO emission is used to estimate the outflow mass, and no
correction is made for a velocity-dependent opacity,
the resultant mass spectrum
power-law slope will be {\it underestimated}.
This can be shown using our data. If we assume that \cooz is 
optically thin or that it has a constant
(velocity-independent) opacity;
we obtain $\gamma \sim 3.2$ for the HH~300
outflow mass spectrum (uncorrected for ambient emission), while 
the velocity-dependent-opacity-corrected value is $\sim$ 5.2 (see Figure~7b).
But, the fact that the HH~300, B5~IRS~1, 
and Circinus outflows all have large values
of $\gamma$ does not necessarily mean that $\gamma$ 
would dramatically increase for 
all other outflows if they were to be corrected
for velocity-dependent opacity. The values of the ``low-velocity''
$\gamma$'s for some young outflows, such as L1448 (data from 
Bachiller et al.~1990, plot from Bachiller \& Tafalla 1999) and 
NGC~2264G (Lada \& Fich 1996), were calculated from fits to the mass spectra 
only for velocities greater than 7.0 km~s$^{-1}$ from the line core, where the 
assumption that the line is optically thin is not a bad one. Thus,
for those flows we would not expect
the value of $\gamma$ to change significantly if a velocity-dependent-opacity
correction of the $^{12}$CO line were included in the analysis. Ultimately,
we expect that if all mass spectra were corrected for opacity,
one would find a significant spread  in the values of 
$\gamma$ perhaps even without a concentration of values around $\gamma \sim 2$.

There are a number of possible explanations for the relatively large value
 of $\gamma$ in HH~300. In their paper, YBB present a list of factors
that could modify the slope of the mass spectrum.
We offer a similar list of factors here, but we focus 
on the elements that could give rise to a steep slope in HH~300.
 
YBB state that their outflow mass estimates are affected by the
velocity at 
which they choose to define an optically thin limit for $R_{12/13}$,
that is, the velocity at which $R_{12/13}$ is set to the carbon isotopic ratio
(see \S 3.3.2 and Figure~7a). In our case, 
changing the isotopic ratio from 62
(the value assumed in this paper), 
to 89 (the solar ratio) changes the total mass
of the red lobe of the HH 300 outflow by less than 1\%, as this change
only affects the ``high outflow velocity'' ($1.9<v<3.2$~km~s$^{-1}$) masses. 
Using any value for the isotopic
ratio from 50 to 89 leaves the power-law slope of the low-velocity 
($0.7<v<1.9$~km~s$^{-1}$) mass spectrum 
unchanged. The high-velocity 
$\gamma$ increases by a maximum of 4\%. 
Thus, the value of the carbon isotopic ratio does not affect our
results considerably. (If we had used only a single power-law fit, the isotopic
ratio would significantly effect the power-law slope, as in YBB.) 

The outflow orientation with respect to the plane of the sky
is another factor which can modify the mass spectrum slope. As shown 
in Smith et al.~(1997), 
where molecular outflows are modeled as 
entrainment of ambient gas by a bow shock, the value of $\gamma$ 
increases as inclination angle (of the flow axis with respect to the plane of
the sky) decreases. 
Specifically, the results of
Smith et al.~(1997) show one model where decreasing the angle between the
outflow axis and the plane of the sky,
 from 60\arcdeg \/ to 15\arcdeg, raises
$\gamma$ from $\sim 1.2$ to $\sim 1.5$.
Although it is very probable that the axis of the HH~300 outflow is
close to the plane of the sky ($i\lesssim 15\arcdeg$), this
would not be enough, by itself, to explain the high value of $\gamma$
($\sim 4$) in the red lobe of HH~300.

Contamination from cloud emission can also affect the slope of the mass
spectrum. 
If one does not take the ambient cloud emission into consideration
 when estimating the outflow
mass, one will overestimate the outflow mass at low velocities, and thus the 
mass spectrum slope will be overestimated. This was shown in section~3.3.2,
where it can be seen that the value of $\gamma$
before correcting for the ambient cloud emission was 
larger before correction ($\gamma=5.2 \pm 0.1$) than after it 
($\gamma=4.0\pm0.2$). On the other hand, our method for correcting the outflow
mass for ambient cloud emission might slightly underestimate the mass from 
the lowest outflow velocity channels (as discussed in \S 3.3.3). 
Underestimating the low-velocity outflow mass would result in an 
underestimation of the power-law slope. 
In order to check if we made a good correction, we constructed
a mass spectrum of a limited
area around clump R3. We made a power-law fit to only the 
channels between outflow velocities of 1.1 and 1.9~km~s$^{-1}$,  
which have predominantly outflow emission (based on the appearance
of channel maps),
and {\it very little or no ambient cloud emission}. 
The resulting mass spectrum, uncorrected for ambient cloud emission,
has a $\gamma=3.9 \pm 0.2$, consistent with 
the mass spectrum of the whole outflow, corrected for ambient cloud emission,
which is $\gamma = 4.0 \pm 0.2$.
Thus, we are convinced that our correction for ambient
cloud emission is a good one,
and that the steep  
value of $\gamma$ we obtain for the red lobe of the HH~300
 outflow is not due to any residual ambient cloud
emission at low velocities.

We believe 
the most important factor in steepening the slope of the HH~300
mass spectrum is outflow 
evolution. Both the models of Smith et al.~(1997) and Downes \& Ray (1999) 
find that the power-spectrum slope of the mass spectrum changes over time. Both
studies find that $\gamma$ steepens as the outflow ages. 
The ambient material that once was accelerated by  
bow shock entrainment, slows down as time goes by, meaning more slow gas
is accumulated, resulting in a steepening of $\gamma$. The simulated
outflows of Smith et al.~(1997) and Downes \& Ray (1999) are relatively young,
(600 and 300 years, respectively), but their results seem to indicate that
the increasing trend of $\gamma$ with age would 
continue as the outflow evolves in time.

Richer et al.~(2000; hereafter RSCBC) do not find any indication of time evolution of the low-velocity ($v<10$~km~s$^{-1}$) $\gamma$ from outflow from low-luminosity (L$_{bol}<10^3$~L$_{\sun}$) 
sources\footnote{YBSBB show a similar compilation
of $\gamma$ values as a function of outflow dynamical age and as a function
of outflow source luminosity. Their compilation is slightly larger than that of RSCBC and includes
the B5-IRS~1 and MMS~9 outflows. Unlike RSCBC, 
YBSBB only use the steeper value of $\gamma$
(which in most outflows is the ``high-velocity $\gamma$'')
for sources that have broken power-law. Similar to RSCBC,
YBSBB's small sample makes it hard to state any definitive conclusion about
outflow evolution.}.
The lack of a trend in their results might be explicable in two ways.
First, RSCBC estimate the age of the outflows using their dynamical age. The dynamical age of an outflow
is estimated using
an average or typical velocity for the source being studied 
and dividing that velocity by the extent of the outflow lobe. 
This is known to be a poor estimate of the age of an outflow
(it could be off by as much as a factor of 10 to 50) as it 
assumes that the outflow has kept the same ``typical'' 
velocity throughout its life and also because it assumes
that all outflowing material comes from the immediate vicinity of
the source (Masson \& Chernin 1993).
Second, the outflow masses are obtained without correcting
for a velocity-dependent opacity in the CO line,
so there could be an underestimation
of the value of $\gamma$ for some of the outflows in RSCBC's
sample. 

If our physical reasoning for the steepening 
of $\gamma$ is correct, we would then
expect that $\gamma$ would evolve differently 
for outflows in physically different environments.
Outflows in much denser mediums will decelerate 
more rapidly than outflows
in low density mediums if the entrainment mechanism 
is a momentum-conserving one,
as it probably is (Masson \& Chernin 1993). 
It would be interesting to conduct 
a survey of outflows in similar environments, 
with mass corrected for CO velocity-dependent opacity and with
accurate age estimates, in 
order to test how $\gamma$ evolves over time.   

The analytical
study by Matzner \& McKee (1999) find that 
it is hard to obtain a value of the power-law mass spectrum slope
very different from 2. This value of $\gamma \sim 2$
is independent of the time history of the driving wind, including its
momentum input as a function of time, and applies to a wide variety of ambient
density distributions. The entrainment mechanism of the ambient medium in the 
Matzner \&  McKee (1999) study does not specifically involve
a bow shock, instead a collimated wind sweeps the 
ambient gas into a momentum-conserving shell, 
following the models of Shu et al.~(1991)
and Li \& Shu (1996).   
The Matzner \& McKee (1999) model does not take into 
account an episodic, precessing outflow which could be modifying
the underlying density and velocity distribution of the surrounding gas
with each mass ejection episode.
As an episodic and precessing outflow changes direction,
different mass outflow ejection episodes may be able to
entrain new ambient material that
previous ejections were not able to effect, as well as to reaccelerate
gas that had already been put in to motion by a previous ejection.
These processes
will certainly effect the mass spectrum slope of an outflow over time.
More on this will be discussed in a subsequent paper (Arce \& Goodman 2001).

\subsection{The episodic and precessing nature of the flow}

The most striking thing about the morphology of the HH~300 outflow is its clumpy
structure. The velocity maps in Figures 4 and 5 show discrete clumps in
space and velocity. It is also notable that the outflow lobe axis  does not 
have a fixed position angle in the plane
of the sky. By drawing a line from the source 
to the maximum emission point for each
of the 5 (red-shifted) clumps in the outflow,
we obtain position angles in the plane of the sky
ranging from 250\arcdeg \/
to 221\arcdeg (see Table~2). 
One unlikely interpretation of the flow's appearance is that 
the underlying wind that drives the outflow
is a wide-angle wind, with an axis very close to 
the plane of the sky. In this interpretation most of the wind
is not detected as the slow radial velocities 
are hidden under the ambient cloud 
emission. The redshifted clumps we detect could then be where the wide angle-wind
is interacting with pre-existing cloud clumps, which have been moved to those 
redshifted velocities by the wind. This scenario seems untenable
for several reasons. First, in a momentum-conserving
wind-clump interaction, the lightest clumps should be
the fastest moving clumps.
This is not the case for the HH~300 clumps, where, for example,
the heaviest clump (R2)
has greater velocities than other, much lighter clumps.
Second, if bow-shock shapes are caused by the wind-clump interaction,
one would expect them to have ``downstream'' wings.
On the contrary, the several clumps that have bow shock 
like structures have all wings pointing back to the source. 

A more plausible explanation is that HH~300 is a 
precessing and episodic outflow, and that the outflow clumps are made
of swept-up ambient gas. The density of the ambient gas in
the B18w region should be on 
the order of $5 \times 10^3$~cm$^{-3}$, and each outflow clump has 
a mass less than 0.5~M$_{\sun}$ (see Table~2).
In order to form a clump
of about 0.5~M$_{\sun}$ by sweeping up gas with
density of $5 \times 10^3$~cm$^{-3}$
a volume of about $4\times10^{52}$~cm$^3$ is needed. 
A cube of side 0.12~pc (or 3\arcmin \/ at a distance of 140~pc)
is more than enough to account for such volume. From the integrated intensity
maps (Figures 4, 5, and 6) we see that the clumps have sizes comparable
to or are bigger than 0.12~pc. 
So, it is realistic to say that most of the mass in the clumps could come from 
gas that has been accelerated very close to the current position of the clump.
Another possible scenario is that the clumps formed (by swept-up ambient gas)
at a position closer to IRAS~04239+2436 and then moved to their 
current position (see \S 4.5). 

Position-velocity (PV) diagrams of HH~300, 
constructed by summing all spectra along an axis with a position angle
of 220\arcdeg \/ (which hereafter will be called the outflow axis)
are shown for the \coto
and \cooz lines in Figures~9 and 10.
It can be seen that at the position of all the
clumps, except R5, there is an increase in the extreme emission velocity.
The velocity structure at the position of each of these clumps is 
characteristic of the prompt entrainment mechanism, which will produce 
the highest velocities at the ``hot spots'' (or local CO maximum of clump)
and decreasing velocity trend towards the source (Bence et al.~1996).
The kinematic behavior where the velocity of an outflow increases
nearly linearly with distance from the outflow source is referred to as the
``Hubble-law'' for outflows. Thus, in our case, we can
 say that there is a different
outflow Hubble-law associated with each clump.
For bow shock models 
(e.g, Zhang \& Zheng~1996; Smith et al.~1997; Downes \& Ray~1999)
the Hubble-law is a natural 
consequence of the fact that the highest velocities are
found at the head of the 
bow shock, while the velocity decreases towards the wings.
Wide-angle wind models of entrainment (Shu et al.~1991; Matzner \& McKee~1999)
only produce a single Hubble-law velocity field.
We can best explain the Hubble-law ``bumps''
at the position of the clumps in Figures~9 and 10
as each being produced by a bow shock at the position of the clumps.

In the bottom panel of Figure~10
we show a plot of the total momentum per unit length (measured in cuts
perpendicular to the outflow axis),
as a function of distance from the source along the flow axis ($z$),
similar to what Chernin \& Masson~(1995)
do for several outflows. It is clear that at the position of
each redshifted clump there is a peak in the momentum. The plot in
the bottom panel
of Figure~10 looks very similar to the plots produced by the precessing
jet model of Cliffe, Frank, \& Jones~(1996).
In their models, each momentum peak 
corresponds to different shock structures in a precessing jet.

In HH~300, from the morphology of the knots in the outflow,
the PV diagram, and the momentum distribution all support a picture
where the outflow is made up of 
several bow shocks along its axis, where each shock corresponds to a 
different ejection event, with a different ejection axis.

Recently, it has been established that it
is common for the ejection axis of outflows to
precess or wander (e.g., RBD).
It is believed that outflow
precession is due to the precession of the outflow source,
which could itself be caused by the tidal interaction between the 
circumstellar disk of the outflow source and one (or many) stellar
companion(s) of the source (e.g., Terquem et al.~1999). Reipurth
et al.~(2000) find that the source of the HH~300 jet is a 
binary with a separation of only 42~AU, which could explain 
the precessing nature of the HH~300 outflow. If
we assume that clumps R1 and R2, the two clumps 
with the most extreme position angles, are close to 
the edges of the precession cone, then the precession
period is approximately twice the difference in eruption age, 
$2(\tau_{dyn,R2}-\tau_{dyn,R1}) \approx 3400$~yr
(see Table~2). Terquem et al.~(1999) give an approximate expression for the
precession period of a disk around a young star in a binary system, where the
disk surrounds only the primary star. The expression is given in terms
of the primary mass, the mass ratio of the two stars, and
the ratio of the disk radius to the 
binary orbit radius.
If we assume a primary mass
of 0.5~M$_{\sun}$, and a primary to secondary mass ratio of about 1.6,
the precession period is about 4000~yr. Hence, it is plausible that the 
precessing nature of the HH~300 outflow could be explained by the fact
that its source lies in a multiple-star system.  

\subsection{Effects on the cloud}

Although the HH~300 outflow is not impressively massive or energetic
(like R Mon, Wolf et al.~1990),
it is still an ``average'' outflow (in the context of the
outflow catalog of Wu et al.~1996), 
which can apparently significantly effect the 
velocity structure of its host cloud.
We have already 
shown that the HH~300 flow is depositing a fair amount of momentum 
over a notable volume of its environment.
Another way of assessing the effect
that an outflow has on the parent cloud is 
by studying the behavior of
the velocity dispersion as a function of position. By
doing so we can examine the effects that the flow has on the cloud kinematics. 
To do this, we fitted a single gaussian
to the \thco spectral component associated with B18w, for all 
spectra southwest of the source. We then
made a contour plot of the velocity width
($\Delta v$) obtained through the gaussian
fit (Figure~11). It appears that the 
outflow clumps are associated
with peaks in the \thco velocity width.  
Many of the \thco spectra at the large 
$\Delta v$ positions exhibit two peaks.
We interpret these two separate velocity components 
as one arising from the
cloud emission, and one arising from both cloud and outflowing \thco emission.
The existence of these multiple peaks implies that in certain positions the
outflow has put substantial amounts of gas in motion in order for the outflow
spectra to appear as a separate velocity component from that of the cloud,
rather than just a low-emission high-velocity wing in the cloud component.
Hence, detecting one (or more) easily identifiable
velocity components attributable to an outflow is yet another 
indication that the flow is having a major  effect on the cloud
gas kinematics.

Different velocity components in outflow spectra from high-density tracers
have been observed in the Mon~R2 outflow (Tafalla, Bachiller \& Wright 1994).
The clumps responsible for the multiple peaks in the CS spectra of the 
Mon~R2 outflow
have been interpreted as gas from the dense core that has been put 
into motion (Tafalla et al.~1997). 
On the other hand, the 
multiple velocity components arising from outflow $^{12}$CO
spectra observed in some low- and medium-mass stars (e.g., 
L1448 [Bachiller et al.~1990];
IRAS~03282+3035 [Bachiller, Mart\'{\i}n-Pintado, \& Planesas ~1991];
HH~111 [Cernicharo \& Reipurth~1996]),  
usually called high-velocity bullets,
have masses of the order
of 10$^{-3}$~M$_{\sun}$ (e.g., Hatchell, Fuller, \& Ladd~1999),
and are thought to 
be composed of jet material. 
The HH~300 redshifted clumps have relatively 
large masses ($\sim 0.1$~M$_{\sun}$, see Table~2),
which makes it probable that they are mainly made up of swept-up material,
more like the high-density clumps of the Mon~R2 outflow.
Therefore, we conclude that 
the $^{13}$CO clumps in HH~300 consist of medium-density
($n \sim 10^{3}$~cm$^{-3}$) gas entrained by the outflow. 

We can quantify the effect that the HH~300 flow has on its parent cloud
using different methods. 

One method is to compare the flow's total energy 
with the cloud's total turbulent kinetic energy. 
The total kinetic energy of a cloud where thermal motions are a negligible
part of $\Delta V$,
is given by $E_{turb}=\frac{3}{16\ln 2} M_{cloud} \Delta V^2$, where
$M_{cloud}$ is the cloud mass and $\Delta V$ is the observed
FWHM velocity line width.
In our case, from the \thco data, the mass of the whole
B18w filamentary structure (seen in Figure~3b) is
$M_{cloud}=72$~M$_{\sun}$.  The velocity width is
$\Delta V \sim 0.8$~km~s$^{-1}$, so $E_{turb} \sim 3 \times 10^{44}$~ergs.
The outflow kinetic energy is about $2.6~(\sin i)^{-2} \times 10^{43}$~ergs,
thus the HH~300 outflow has the potential of making a substantial 
contribution to the turbulent energy of the cloud. For $i=17\arcdeg$, the outflow kinetic energy is equal to the kinetic energy of the B18w region.
Armstrong \& Winnewisser (1989) discovered that the kinetic energy of 
a parsec-scale outflow in L673 could be 6 to 0.4 times (depending on the assumed
projection angle and the amount of mass ``hidden'' under the line core)
the turbulent energy in the cloud. At the time Armstrong \& Winnewisser (1989)
made their study, astronomers had no idea of the common occurrence of 
parsec-scale HH flows from young stellar objects. Now, with the discovery
that HH flows may typically extend for several parsecs (RBD)
we should reconsider the role of outflows in driving the 
turbulent energy of molecular clouds.

We can also asses the importance of the HH~300 outflow
energy input on its parent cloud
by using the results of numerical simulations. 
The self-consistent 
magnetohydrodynamics (MHD) simulations of Gammie \& Ostriker (1996) study 
MHD turbulence under density, temperature and magnetic field conditions
representative of those found in Galactic molecular clouds. The models
use ``slab symmetry'', in which all variables are a function of one
independent spatial variable $x$ and time $t$. Using equation 34
in Gammie \& Ostriker (1996),
we can derive the input power needed to support a 
molecular cloud in equilibrium.
We approximate the  ``linear'' dimension ($L$) of B18w to be twice the geometric
mean of its axes ($\sim 1.1$~pc). We assume 
a mean density of  $5\times10^{3}$~cm$^{-3}$, and
a $\beta$ parameter\footnote{The $\beta$ parameter 
is defined as the ratio of the square of the sound speed to the square of the
Alfv\'en speed in the cloud, that is $\beta =c^{2}_{s}/v^{2}_{A}$} of 0.01.
Using these approximations we find that we would need about 0.4 L$_{\sun}$
of input power at a scale 
of 1~pc (the size of the outflow lobe)
to counter the dissipation of MHD turbulence in the 
B18w region and thus support it against gradual gravitational contraction. 

The power of an outflow is usually estimated by dividing the
outflow kinetic energy by the dynamical age of the outflow.
The conventional way to estimate the dynamical age of a molecular outflow assumes
that all the gas in the outflow originates at the young star. This 
assumption is wrong, since the vast majority of the gas in a molecular outflow
comes from the entrained gas in the host cloud (along the extent of the outflow)
 that has been put in motion by 
the underlying stellar wind. Since there is no way to obtain 
an accurate estimate of the outflow lifetime, we
estimate a lower and an upper bound to the dynamical age in order to estimate
an upper and lower bound to the outflow power.  
We estimate the dynamical age lower limit to be 
the time it has taken HH~300A (the HH knot which currently lies further to 
the source) to travel to its current position. The distance from the source
to HH~300A is 1.2~pc, and if 
we assume a typical HH jet velocity of 200~km~s$^{-1}$, we then
obtain a dynamical age of about 5900~yr. As an upper limit on the age we use
$2 \times 10^{5}$ years, the statistical lifetime of outflows as derived
by the study of Parker, Padman, \& Scott~(1991).
We then estimate the lower and upper limits of the outflow power to be 
$1.1 \times 10^{-3}~(\sin i)^{-2}$ to 
$3.6 \times 10^{-2}~(\sin i)^{-2}$~L$_{\sun}$.
If we assume that $i\sim 10\arcdeg$, 
then the outflow power's lower limit is only about a factor of ten less than
what is needed to drive the MHD turbulence in B18w. Thus, the HH~300 outflow 
has the potential to make a substantial contribution to the power needed to
sustain the MHD turbulence in B18w.

Yet another way to estimate the effects of the flow on the cloud is to compare
the flow's energy with the cloud's binding energy. 
The B18w cloud binding energy ($\sim GM_{cloud}^{2}/R$), where we estimate 
$R$ to be the geometric mean of the short and long axes of B18w, 
is about $8\times 10^{44}$~ergs. Hence, if $i\sim 10\arcdeg$ the HH~300 outflow
would have enough kinetic energy
to potentially disperse a major fraction of the B18w region.
From our estimates (see \S 4.1) 
we see that HH~300 is effecting a notable fraction of B18w, so
whether most of the outflow kinetic energy will be converted to
turbulent energy or it will be used to directly
disperse the cloud gas (see \S 4.6 for
more on this), it is clear that the HH~300 outflow will have a major
effect on the evolution of B18w.

\subsection{The structure of B18w}

The axis of the HH~300 redshifted outflow lobe is parallel to B18w's
long axis, as can be seen in both Figure~2 and Figure~6. 
The B18w region looks like a protuberance that sticks out from the main
B18 molecular cloud (see Figure~1).
Given the nice positional coincidence of the outflow's and the dark
cloud's axes, in addition to the particular shape of B18w and its position
with respect to the rest of the B18 cloud, we
originally flirted with the idea that the HH~300 outflow was creating 
the B18w dark cloud. 
We hypothesized that the outflow was dragging gas and dust
from the star-forming core and forming the B18w dark cloud. Without calculating
any estimates of mass for HH~300 and B18w, this seemed like a reasonable idea.
Similarly, Bence et al.~(1996) speculate that the RNO~43 outflow's
 most southerly bow shock
could be responsible for ejecting material from the main cloud and creating
 a spur of cloud material, seen in $^{12}$CO emission
and in extinction (see their Figure~19). But, once we 
estimated the mass of the outflowing material to be only 7\% that of the
surrounding ambient gas, in addition to studying the kinematics of the gas,
we concluded that what we observe to be the ``current''
HH~300 molecular outflow has not created the B18w region.
Moreover, if this scenario were true one would expect the blue and the red
lobes to drag along similar amounts of gas. As seen in Figures 1 and 8, the 
B18w cloud extends southwest with respect to the outflow source
but does not extend northeast of the source, where one would expect the blue 
lobe to be. Such a high asymmetry in the ``dragging process'' is highly 
unlikely.
Even if this scenario does not apply to the HH~300 outflow,
we still believe that the process in which an outflow ejects and/or drags
gas and dust from a dense core (or any other dense region in the parent cloud)
and moves it to distances of the order
of 0.1 to 1~pc could take place in other sources.
If HH~300 is not ``producing'' the B18w region, then
the fact that B18w and HH~300 have long axes
with similar position angles is a coincidence.  
In that case, HH~300 has a shape similar to B18w because B18w 
is the cloud that ``supplies'' the gas that HH~300 is made of. 
That is, the parent cloud is constraining the morphology of the {\it observed}
outflow.

\subsection{The cumulative effect of outflows}

Our observations clearly show that even low-mass young stellar objects 
with a luminosity of about 1 L$_{\sun}$ can affect their surrounding density 
and kinematical distribution on parsec scales. From our results, we 
can speculate that the cumulative action of many outflows, from different 
YSOs could have a profound effect on a cloud's evolution and fate.
The cumulative effect of many outflows could  be:
1) to produce just enough turbulent energy to support
the cloud against gravitational collapse;  2) to greatly increase the turbulent
energy to more than the binding energy of the cloud; or 3)
to put enough ambient gas into motion as to directly disperse the cloud.
If enough
gas is moved out of the gravitational potential well of the cloud,
the cloud would eventually disperse.
If the ambient density is relatively high and 
the outflow momentum is relatively low, then the outflowing 
material will eventually slow down 
through the interaction with the ambient medium, mixing with
its surroundings, and feeding the turbulent energy of the cloud.
On the other hand,
if the ambient density is relatively low and the outflow momentum is high, the 
swept-up gas will barely decelerate and it could eventually be ``pushed away'' 
from the cloud. It is hard to conclude which of these two scenarios will
prevail in a given cloud, without modeling the interface between outflow
and cloud. Further theoretical and numerical investigations should concentrate
on how efficiently bulk motions, produced by outflows, can decay to produce
turbulence as a function of a cloud's density and outflow energy. 
Now that it has been established that individual outflows are energetic
enough, and large enough to substantially affect a big portion of 
their host cloud, observational studies should concentrate in investigating 
the cumulative effects that many outflows have on a
molecular cloud and on a cloud complex (e.g., see BRLB).

\section{Conclusion}

We mapped the red lobe of the giant molecular outflow
associated with the HH~300 flow in the \coto line, with a 
beam size of 27\arcsec. We also made a more extended map 
of the gas surrounding the HH~300 flow at the \cooz and 
\thco lines, with  45\arcsec \/ and 47\arcsec \/ beam sizes, respectively.
By observing a large extent of the gas surrounding the
outflow we are able to study the outflow in the context of
its surrounding medium. Also, the \thco observations help us 
assess the effects the outflow has on the surrounding
moderate-density ($n \sim 10^3$~cm$^{-3}$) gas structure and kinematics. 
In addition, the combined \cooz and \thco line observations
enable us to estimate the mass of the outflow
by correcting for the velocity-dependent opacity
of the \cooz line. Due to ``contamination''
from emission of another molecular cloud along the same light-of-sight
we are not able to study the blue lobe of the HH~300 outflow.

Our results show that the HH~300 outflow has a very clumpy structure.
We identify five $^{12}$CO redshifted clumps, each of which is readily
apparent in position-position-velocity space. They each 
have masses of the order of a few $10^{-1}$~M$_{\sun}$ and reach
radial velocities of about 3~km~s$^{-1}$  from the ambient cloud velocity.
Given the low inclination of the outflow with respect to the plane of the
sky, the deprojected velocities of these clumps are likely to be higher than  
15~km~s$^{-1}$. Such high masses and low velocities suggest that these
clumps are made of swept-up ambient gas.
The spatial distribution of the clumps, the velocity structure, and
the momentum distribution of the outflow 
indicate that these clumps arise from prompt entrainment, 
most probably produced by bow shocks, arising from the young star's
mass ejection episodes.
Each of these clumps has a different orientation on the
plane of the sky (or position angle) with respect to the outflow source.
We conclude that HH~300 is a precessing and episodic outflow.

We obtain a power-law mass spectrum slope of $-4.0\pm0.2$ for low
outflow velocities and a slope of $-7.8\pm0.4$ for high outflow
velocities, for the red lobe  of HH~300.
These slopes are steeper than the average for other outflows.
To obtain the outflow mass we used a velocity-dependent
\cooz optical depth, and we
corrected the low-velocity outflow mass for ``contaminating'' ambient cloud
emission. Previous outflow studies, with the exception of Bally et al.~(1999),
Yu et al.~(1999), Yu et al.~(2000)
from which we obtain the idea, do not correct 
for the velocity-dependent opacity of the line, when obtaining 
outflow parameters from their $^{12}$CO data. As we have shown, not applying
this correction will underestimate the low-velocity outflow mass.
We believe that some low-velocity outflows will exhibit a steeper mass spectrum
power-law slope ($\gamma$)
once their masses have been corrected for the velocity-dependence
of the $^{12}$CO line. Thus, we expect that for a sample of outflows
(with different ages) the value of $\gamma$ will have a
wider range of values than what is currently found in the literature.
The fact that HH~300 has a steep mass 
spectrum power-law slope ($\gamma \sim 4$) is most probably due to the 
evolution of outflow mass kinematics. 
Ambient material accelerated by several episodes of mass ejection by the 
young star will eventually slow down, leading to 
accumulation of slow gas, and a steepening of $\gamma$.

The HH~300 outflow, although not extremely powerful, 
is depositing a fair amount of momentum 
($3.2~(\sin i)^{-1}$~M$_{\sun}$~km~s$^{-1}$) and kinetic energy
($2.6~(\sin i)^{-2} \times 10^{43}$~ergs)
over a notable volume ($\sim 11$\%) of its parent dark cloud (B18w). 
We qualify the effects that the HH~300 flow has on 
its host cloud by using  three different methods.
If HH~300 has an inclination to the plane of the sky ($i$)
of about 10\arcdeg, then the red lobe of the HH300 molecular 
outflow has more than enough energy to
supply the turbulent energy of B18w, more than 
enough energy to gravitationally unbind the 18w region, and enough 
power to make a substantial contribution to 
the power needed to sustain the MHD turbulence in B18w. 
In addition, the $^{13}$CO data indicate that the HH~300 flow is having 
a major effect on the cloud gas kinematics, and is able to redistribute
considerable amounts of its surrounding medium-density 
($\sim 10^3$~cm$^{-3}$) gas.
 
Our study clearly shows that
even low-mass young stellar objects of about 1~L$_{\sun}$ can produce drastic
changes in their surrounding environment, and affect a notable volume of
their parent clouds.
This is probably 
a characteristic of giant HH flows, which typically extend over a few parsecs
on the sky, and precess. 
Giant outflows have the potential
of modifying the velocity and density structures not only of their host
core, but also of their host cloud, even at parsec-scale distances from the
source. The cumulative action of giant outflows from young stellar objects 
will certainly have a profound effect on a cloud's evolution and fate.

\acknowledgements

We are grateful to Bo Reipurth, John Bally and Ka Chun Yu
for their useful comments, and to the
National Science Foundation (AST-9457456 and AST-9721455)
for supporting our efforts.

%%%%Figure Section%%%%
\clearpage
%Figure 1
\figcaption[figure1.eps]{\thco integrated intensity map of the B18 cloud
in Taurus for the velocity range $2.9 < v < 8.9$~km~s$^{-1}$. The map is
part of the large scale \thco map of the Taurus molecular cloud complex
from Mizuno et al.~(1995). The source of
the HH~300 flow, IRAS~04239+2436, is shown with a white star symbol.
The integrated intensity map of the HH~300 molecular outflow's red lobe
is shown in white contours. The B18 and B18w regions are identified.}

%Figure Two
\figcaption[figure2.eps]{({\it Top}) Average \cooz and \thco 
spectra of the area mapped in $^{12}$CO(2--1),
which includes most of the HH~300 outflow red lobe and
some cloud regions unaffected by the outflow. ({\it Bottom})
Average \coto spectrum over the same area as the spectra above (the area
covered by the \coto map).}

%Figure Three
\figcaption[figure3.eps]{Velocity-range-integrated
 intensity maps of the \thco emission. The
velocity interval of integration is given on the top of each panel. The
first contour and the contour steps are given in parenthesis on the bottom
left corner of each panel in units of K~km~s$^{-1}$.
The position of the outflow source, IRAS~04239+2436, is identified
with a star symbol.
({\it a}) Emission from cloud A. 
({\it b}) Emission from B18w. ({\it c}) Emission from B18w ambient gas
plus HH~300 outflow emission. ({\it d}) Emission southwest of the
outflow source is solely from HH~300 outflow. The emission northeast
of the source is from another cloud on the same line of sight.
({\it e}) Emission southeast of outflow source is from
the highest velocity outflowing \thco gas.}

%Figure Four
\figcaption[figure4.eps]{Velocity-range-integrated intensity
maps of the \coto emission
from the redshifted lobe of the HH~300 molecular outflow. The
velocity interval of integration is given on the bottom of each panel.
The first contour and the contour steps are 0.325~K~km~s$^{-1}$ for all maps. 
The position of the outflow source, IRAS~04239+2436, is identified with a star symbol. The five outflow clumps R1, R2, R3, R4, and R5 are labeled.}

%Figure Five
\figcaption[figure5.eps]{Velocity-range-integrated intensity
maps of the \cooz emission
from the redshifted lobe of the HH~300 molecular outflow. The
velocity interval of integration is given on the bottom of each panel. 
The first contour and the contour steps are 0.22~K~km~s$^{-1}$ for all maps.
The position of the outflow source, IRAS~04239+2436, is identified with a star symbol. The five outflow clumps R1, R2, R3, R4, and R5 are labeled.}

%Figure Six
\figcaption[figure6.eps]{Integrated intensity contour map of the \cooz emission
over the range $7.53<v<9.94$~km~s$^{-1}$,
superimposed on a wide-field H$\alpha$+[S~II] (optical)
CCD image from Reipurth, Bally, \& Devine (1997). The starting contour of the
integrated intensity map is 0.55~K~km~s$^{-1}$, and the contour steps
are 0.275~K~km~s$^{-1}$. The A, B, C, and D
HH knot groups from the HH~300 flow are labeled. We also label the molecular
outflow clumps R1, R2, R3, R4, and R5. The location of the deeply embedded
outflow source, IRAS~04239+2436, is shown. The B18w cloud is seen in the 
optical image as a region with no background stars (due to extinction).}

%Figure Seven 
\figcaption[figure7.eps]{({\it a}) Main beam
temperature (or intensity) ratio of \cooz to $^{13}$CO(1--0),
also denoted in the text as $R_{12/13}$, as a function of 
observed velocity ($v$).
 The filled circle symbols are the points
used for the second-order polynomial fit to
$R_{12/13}$. The solid line is the resulting fit.
({\it b}) Mass spectrum, or mass in a 0.22~km~s$^{-1}$-wide channel 
as a function of outflow velocity ($v_{out}=v-V_{LSR}$),
 for the red lobe of the HH~300 molecular
outflow. The squares denote the mass obtained assuming that \cooz is 
optically thin (Mass $\alpha$ $\int T_{mb} dv$). 
The diamonds represent the mass obtained assuming that the \cooz line
opacity is constant ($R_{12/13} =13$).
The triangles denote 
the mass assuming that $R_{12/13}$, and thus the opacity of the 
\cooz line, varies as a function of velocity, as shown in Figure~7a.
For each of these three different mass estimates we show two power-law
fits, one for low outflow velocities
 between 0.75 and 1.85~km~s$^{-1}$,
and another for high outflow velocities between 1.85 and 3.17~km~s$^{-1}$.
The points at $v_{out}=0.53$~km~s$^{-1}$ are real, but are excluded
from the fits, as they lie too close to the ambient line core. 
Velocities are {\it not} corrected for outflow inclination.
The long and short dashed lines are power-law fits to the optically thin
mass estimate and the constant opacity mass estimate,
respectively. 
The solid lines show power-law fits to the velocity-dependent opacity
corrected mass spectrum.
The power-law exponents from the fits are as follows.
$\tau << 1$: $-3.3 \pm 0.1$ (low velocity); $-6.1 \pm 0.2$ (high velocity).
$\tau=$~constant: $-3.3 \pm 0.1$ (low velocity); $-6.1 \pm 0.2$ (high velocity).
$\tau(v)$: $-5.2\pm 0.1$ (low velocity);  $-7.8\pm0.4$ (high velocity).
({\it c}) Mass in a 0.22~km~s$^{-1}$-wide channel, as a function of observed
velocity. Solid line is a two-gaussian fit to the filled triangle symbols
(points that are not ``contaminated'' by emission from the HH~300 outflow).
({\it d}) Mass spectrum corrected for ambient cloud emission.
The mass is corrected
for ambient cloud emission by subtracting the modeled ambient cloud
mass (given by the gaussian fit in Figure~7c) from the velocity-dependent
opacity corrected mass (shown in Figure~7b in triangle symbols).
See discussion in text (\S 3.3.3).
Similar to Figure~7b, the resultant power-law fits 
are shown by solid lines. The slopes of the fits are $-4.0 \pm 0.2$ 
and $-7.8 \pm 0.4$.}

%Figure Eight
\figcaption[figure8.eps]{Contour map of outflow total momentum, {\it not}
corrected for inclination angle, for
outflow velocities greater than 0.97~km~s$^{-1}$. The momentum contour map is
superimposed on a B18w grey-scale map of \thco integrated intensity for $5.8<v<7.1$~km~s$^{-1}$, as shown in Figure~3b.
The momentum was obtained using the velocity-dependent opacity 
correction technique described in the text. The starting contour and the
contour steps on the momentum map are both 
$1.5\times 10^{-4}$~M$_{\sun}$~km~s$^{-1}$.}

%Figure Nine
\figcaption[figure9.eps]{{\it Top}. \coto integrated intensity contour
map for velocities between 7.30 and 10.55~km~s$^{-1}$. 
{\it Bottom}. \coto Position-Velocity diagram of the redshifted lobe of HH~300.
The outflow velocities, as well as the ambient gas velocities
of the B18w cloud and of cloud A, are indicated on the right
 of the bottom panel.
The position of each clump along the outflow axis is marked with a vertical 
solid line. Notice how at the position of each molecular outflow clump
there is an increase in the velocity. The increase in velocity 
at most of the clump positions, as shown
with the dotted curved lines, follows a ``Hubble-law'',
where the velocity increases proportionally to the distance from the source. 
This velocity structure is consistent with 
the outflow clumps being produced by bow shock prompt entrainment from
an underlying wind (see \S 4.3 for discussion).}

%Figure Ten
\figcaption[figure10.eps]{{\it Top}. \cooz integrated intensity contour
map, as shown in Figure~6. {\it Middle}. \cooz Position-Velocity diagram of the 
redshifted lobe of HH~300. {\it Bottom}. Distribution of the outflow
momentum per unit length ($dM/dz$) along the HH~300 outflow axis. The triangle
symbol represents the total momentum over the width of the outflow lobe.
The outflow velocities, as well as the ambient gas velocities
of the B18w cloud and of cloud A, are indicated on the right
of the middle panel.
The position of each clump along the outflow axis is marked with a vertical 
solid line. Notice that at the positions of the R1, R2, and R3 clumps,
the 3 most prominent clumps in
\cooz, there is an increase in velocity and momentum 
(see \S 4.3 for discussion).}

%Figure Eleven
\figcaption[figure11.eps]{\cooz integrated intensity map of the 
HH~300 outflow, as in Figure~6, superimposed on a grey-scale map
of \thco line width (FWHM).
First grey tone corresponds to a width of 1.0~km~s$^{-1}$.
Between each grey tone there is an increment of 0.2~km~s$^{-1}$.  
Coordinate offsets are given 
with respect to the outflow source's position at
($\alpha$,$\delta$)$_{1950}$ = ($4^h23^m54^s.5$, 24\arcdeg36\arcmin54\arcsec). 
Six example \thco 
spectra at different position are shown.
In all spectra, the cloud A and the B18w spectral components can be seen.
The B18w (+ HH~300 in some cases) spectral component is highlighted in grey.
Three spectra are from regions
presumably unaffected by the HH~300 outflow, and three are from regions of the
cloud affected by the outflow. On these last three, 
a vertical solid line indicates the position of
the separate spectral component, presumably due to the outflow. 
All spectra are plotted with the same axes ranges: 0 to 12~km~s$^{-1}$
in the velocity axis; and -0.2 to 3.2~K on the antenna temperature axis.}

\clearpage

\begin{deluxetable}{crrrrr}
\tablecolumns{6}
\tablecaption{Detectability of Clumps in HH~300 Redshifted Lobe
\label{clumpobsinfo}}
\tablehead{
\colhead {} &
\multicolumn{3}{c}{Integrated Intensity Maps\tablenotemark{a}} &
\multicolumn{2}{c}{Position-Velocity Diagrams\tablenotemark{b}}
\\ \cline{2-4}   \cline{5-6}
\colhead{Clump} & 
\colhead{\thco} & \colhead{\cooz} & \colhead{\coto} &
\colhead{\cooz} & \colhead{\coto}
}
\startdata

R1 & yes & yes & yes & yes & yes\nl
R2 & yes & yes & yes & yes & yes\nl
R3 & yes\tablenotemark{c} & yes & yes & yes & yes\nl
R4 & no & $\times$\tablenotemark{d} & yes & $\times$\tablenotemark{e} & yes\nl
R5 & no & yes & yes\tablenotemark{f} & no & yes\enddata  
\tablenotetext{a}{Here we note if the clump structure can be seen in any of the
integrated intensity velocity maps of the different molecular lines 
(Figures 3, 4 and 5)}
\tablenotetext{b}{Here we note if the clump shows a velocity increase in the
the position-velocity (PV) diagrams (Figures~9 and 10).}
\tablenotetext{c}{The R3 clump is seen as two clumps in $^{13}$CO(1--0).}
\tablenotetext{d}{There is \cooz emission at the position of clump R4, but the
\cooz maps do not show the same, clearly defined, clump structure as in
the \coto line.}
\tablenotetext{e}{There is not a clear peak in the velocity at the position
of clump R4 in the  \cooz PV diagram, though there is high-velocity gas
(see Figure~10).}
\tablenotetext{f}{Clump R5 can only be seen partly in the \coto data
because of the limited extent of the \coto map (see Figure~4a).}

\end{deluxetable}

\clearpage

\begin{deluxetable}{cccrc}
\tablecolumns{5}
\tablecaption{Physical Properties of Clumps in HH~300 Redshift Lobe
\label{clumpinfo}}
\tablehead{
\colhead{Clump} & \colhead{d\tablenotemark{a} \/ [pc]} 
& \colhead{Angle\tablenotemark{b} \/ [deg]} &
\colhead{Age\tablenotemark{c} \/ [yr]} 
& \colhead{Mass\tablenotemark{d} \/ [M$_{\sun}$]}
}
\startdata
R1 & 0.16 & 250 &  800 & 0.07\nl
R2 & 0.53 & 221 & 2500 & 0.21\nl
R3 & 0.81 & 228 & 4000 & 0.23\tablenotemark{e}\nl
R4 & 0.98 & 225 & 4800 & ---\tablenotemark{e}\nl
R5 & 1.03 & 238 & 5000 & 0.02\tablenotemark{f}\enddata
\tablenotetext{a}{Distance from the outflow source to the maximum emission
point on clump}
\tablenotetext{b}{Position angle (east of north) of line intersecting
outflow source and maximum emission point on clump.}
\tablenotetext{c}{Dynamical age of the clump assuming mass ejection from source,
responsible for the clump, has a tangential velocity of 200~km~s$^{-1}$.}
\tablenotetext{d}{Mass of clump for $v_{out}\geq 0.97$~km~s$^{-1}$, uncorrected
for cloud emission. Emission at $v_{out}\geq 0.97$~km~s$^{-1}$ is dominated by
outflow clump emission, cloud emission dominates at slower velocity channels.}
\tablenotetext{e}{Clump R3 and R4 are indistinguishable (unresolved)
in \cooz and \thco lines, the total mass of both is given in the R3 mass entry.}
\tablenotetext{f}{Unlike the other clumps,
clump R5 is seen better at lower velocities, $v_{out} \sim 0.75$~km~s$^{-1}$.
Mass of clump R5 for $v_{out}\geq 0.75$~km~s$^{-1}$ is 0.11~M$_{\sun}$.}

\end{deluxetable}

\clearpage

\begin{deluxetable}{lcc}
\tablecolumns{3}
\tablecaption{Mass, Momentum, and Kinetic Energy of the Redshifted
Lobe of HH~300 
\label{energetic}}
\tablehead{
\colhead {} & \colhead{$v_{out} \geq 0.75$~km~s$^{-1}$} &
\colhead{$v_{out} \geq 0.53$~km~s$^{-1}$}
}
\startdata
Mass & 2.4 & 4.3~M$_{\sun}$\nl
Momentum & $2.2~(\sin i)^{-1}$ & $3.2~(\sin i)^{-1}$~M$_{\sun}$~km~s$^{-1}$\nl
Kinetic Energy & $2.1~(\sin i)^{-2}$ &
   $2.6~(\sin i)^{-2} \times 10^{43}$~ergs\enddata

\end{deluxetable}

\clearpage

\begin{deluxetable}{lccc}
\small
\tablecolumns{4}
\tablecaption{Comparison of $\gamma$ for different velocity-dependent 
opacity corrected outflows 
\label{gammacomp}}
\tablehead{
\colhead {} &
\multicolumn{2}{c}{Average $\gamma$\tablenotemark{a}} & 
\colhead{}
\\ \cline{2-3}   
\colhead{Outflow} & 
\colhead{Without ambient cloud correction} & \colhead{With ambient cloud
subtracted} & \colhead{References}
}
\startdata

Circinus Flow A & 4.0 & ---\tablenotemark{b} & c\nl
Circinus Flow B & 3.7 & ---\tablenotemark{b} & c\nl
Circinus Flow C & 3.8 & ---\tablenotemark{b} & c\nl
B5-IRS~1 & 3.9 & 2.7 & d\nl
MMS~9 & 3.8 & 1.8 & e\nl
HH~300 & 5.2 & 4.0 & f\enddata
\tablenotetext{a}{The number given is the average value of $\gamma$,
over both lobes of the given outflow (except for HH~300,
where only one lobe was studied). The value of $\gamma$ is defined to be the
power-law slope in the relation $dM(v)/dv ~ \alpha ~ v^{-\gamma}$.
For the two outflows with mass spectra exhibiting power-laws
(HH~300 and MMS~9), we show the ``low-velocity'' $\gamma$.}
\tablenotetext{b}{There is no mass spectrum with ambient cloud subtracted
for these outflows.}
\tablenotetext{c}{Bally et al.~1999}
\tablenotetext{d}{Yu et al.~1999}
\tablenotetext{e}{Yu et al.~2000}
\tablenotetext{f}{This paper}

\end{deluxetable}

\end{document}